# Proposed optimal LSP selection method in MPLS networks

## Shin-ichi Kuribayashi[1]

[1]Department of Computer and Information Science, Seikei University, Japan

E-mail: kuribayashi@.st.seikei.ac.jp

### Abstract

*Multi-Protocol Label Switching (MPLS) had been deployed by many data networking service providers, including the next-generation mobile backhaul networks, because of its undeniable potential in terms of virtual private network (VPN) management, traffic engineering, etc. In MPLS networks, IP packets are transmitted along a Label Switched Path (LSP) established between edge nodes. To improve the efficiency of resource use in MPLS networks, it is essential to utilize the LSPs efficiently.*

*This paper proposes a method of selecting the optimal LSP pair from among multiple LSP pairs which are established between the same pair of edge nodes, on the assumption that both the upward and downward LSPs are established as a pair (both-way operation). It is supposed that both upward and downward bandwidths are allocated simultaneously in the selected LSP pair for each service request. It is demonstrated by simulation evaluations that the proposal method could reduce the total amount of the bandwidth required by up to 15% compared with the conventional selection method. The proposed method can also reuse the know-how and management tools in many existing networks which are based on both-way operation.*

## 1. Introduction

The undeniable potential of Multi-Protocol Label Switching (MPLS) [1],[2] in terms of virtual private network (VPN) management, traffic engineering, path protection, and rapid recovery from network failures, has resulted in most major data networking service providers either having already deployed MPLS or being in the process of deploying it. A number of data networking service providers have also announced their plans to completely migrate all their services, such as Ethernet, frame relay, and ATM, onto MPLS networks [3]-[14]. Transport-MPLS [15] based on MPLS technology had also been discussed and standardized as a new transmission means in ITU-T. MPLS will be also adopted for the next-generation mobile backhaul networks based on Long Term Evolution (LTE) technology [16]. Therefore, MPLS could be a promising candidate for the core switching infrastructure for all services.

In MPLS networks, IP packets are transmitted along a Label Switched Path (LSP) established between an ingress Label Switched Router (LSR) and an egress LSR. To improve the efficiency of resource use in MPLS networks, it is essential to utilize LSPs effectively. Multiple LSPs will be established between the same pair of ingress LSR and egress LSR, for reasons of reliability, load-balancing, congestion suppression, etc. One example is a situation where spare LSPs are established beforehand in order to avoid interruption of communication if a problem occurs with one LSP. Other examples are the cases where the required bandwidth cannot be secured on a route and so multiple LSPs are established via different routes, or a multipass configuration is adopted to improve resource efficiency [17],[18]. An additional example is use of the link aggregation service [19] with Ethernet emulation services over an MPLS network.

Although an LSP is basically a 'unidirectional' (one-way) connection, in practice it is useful to handle both upward LSP and downward LSP between the same edge nodes as a pair (both-way operation) and to establish them simultaneously. As the design and management of many of existing networks are based on both-way operation, it becomes possible to reuse the know-how and management tools in the existing network. The both-way operation of LSP has already been adopted for transport-MPLS [15]. We call such a pair an '**LSP pair**' hereafter.

This paper proposes a method of selecting the optimal LSP pair from among multiple possible LSP pairs between the same pair of edge nodes. The objective of this paper is to minimize the total amount of bandwidth, not to minimize the processing time or the response time. First, an LSP pair selection method is proposed, on the assumption that both upward and downward bandwidth are allocated simultaneously in the selected LSP pair, per service request. Next, the paper proposes an extension of the LSP pair selection method in which the network delay time is considered. Then the effectiveness of the proposed methods is demonstrated by simulation evaluations. It should be noted that the method proposed here can also be applied to LSP selection in GMPLS (Generalized MPLS) networks or the link selection in link aggregation configuration of Ethernet. This paper is an extension of the study in [20].

## 2. Related works

The following papers are related to research on selecting the best LSP from multiple unidirectional LSPs between the same pair of edge nodes. References [17] proposes a heuristic load-balancing algorithm using the traffic characteristics on the assumption that the load is distributed at the flow level. Reference [21] proposed an automatic bandwidth setting method based on a percentile basis since the existing methods tend to assign more bandwidth in the event of the sudden traffic increase. Furthermore, although it is not MPLS itself, reference [22] proposes a circuit allocation method which classifies flows into priority flows and non-priority flows and fixes the circuits for non-priority flows on the assumption that a link aggregation service is applied. In addition, the authors have proposed a method of selecting one optimal VP among multiple possible bi-directional VPs in an ATM network [23]. However, the above MPLS studies are based on unidirectional LSPs and differ from the model treated in this paper, in which both the upward and downward LSPs are established as a pair (both-way operation), and both upward and downward bandwidths are allocated in the same LSP pair simultaneously.

## 3. Proposal of optimal LSP pair selection method

An actual MPLS network is modeled as in Figure 1 in this paper. That is, there are multiple LSP pairs (n pairs) between the same pair of edge nodes (edge node X and edge node Y in Figure 1). Each LSP pair is established independently via an arbitrary route. Each LSP pair has a different maximum size of upward and backward bandwidth, and a different network delay time, respectively. The maximum size of upward bandwidth is not necessarily the same as that of backward bandwidth in each LSP pair.

It is supposed that each service request requires both upward and downward bandwidths between edge nodes. If there is not enough bandwidth available for both directions in any possible LSP pair, the service request will be rejected. Moreover, it is assumed that the size of the required upward bandwidth is not necessarily the same as that of the downward bandwidth.

This section proposes an algorithm of selecting the optimal LSP pair from among multiple LSP pairs between the same pair of edge nodes. The objective of this paper is to minimize the total amount of bandwidth, not to minimize the processing time or the response time. The following three methods are considered here. Method A is one of conventional methods and

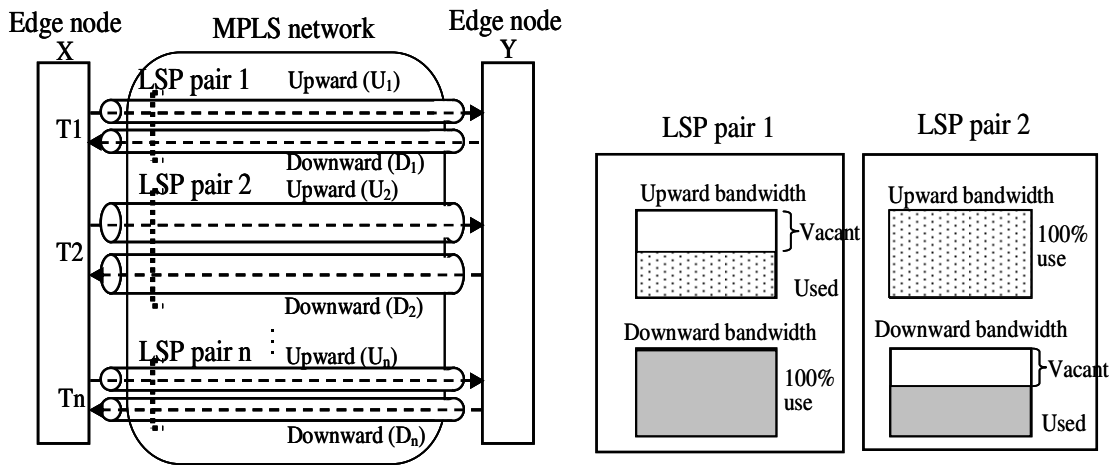

· LSP pair j : a pair of upward LSP and downward LSP between edge node X and edge node Y
· $U_j$ : Maximum upward bandwidth of LSP pair j
  $D_j$ : Maximum downward bandwidth of LSP pair j
· $T_j$ : Network delay time of LSP pair j

Figure 1. MPLS network model for LSP pair selection

*It is not possible to support a request even though spare bandwidth for both directions are available (not available in the same LSP pair).*

Figure 2. Example of 'deadlock state'

Methods B and C are newly proposed in this paper. Methods A and B do not consider the network delay time of each LSP in the case of LSP pair selection. Method C considers the network delay time of each LSP in the case of LSP pair selection.

**[Method A]**

This is one of the conventional methods, which does not consider the situation of both upward bandwidth and downward bandwidth in LSP pair selection. One LSP pair from among n LSP pairs is selected in a round-robin fashion. If there are not enough bandwidth available for the selected LSP pair, the next LSP pair in the pre-defined order will be selected. For each service request, the LSP pair which is next in the pre-defined order will be checked first, regardless of which LSP pair was selected by the previous request. Both upward and downward bandwidths are allocated simultaneously in the selected LSP pair. Note that the request will be rejected if none of the LSP pairs have enough bandwidth available.

**[Method B]**

This method aims to reserve as much bandwidth as possible for future requests that may require a larger size of bandwidth, and to reduce the possibility that the 'deadlock situation' shown in Figure 2 will occur, in which it is not possible to support a request even though spare bandwidth for both directions are available (not available in the same LSP pair).

In Method B, the direction that requires the largest proportionate size of bandwidth is first identified as "**key direction**", comparing the size of required bandwidth with the maximum size of bandwidth for each direction. Then the LSP pair with least available bandwidth of the key direction from among n LSP pairs is selected. Bandwidth allocation algorithm is as follows:

i) Identification of key direction

If $X_u \geqq X_d$ then upward direction is identified as the key direction. Else downward direction is identified as the key direction.

where

$X_u$= {the size of required bandwidth for upward direction}/$X_{u0}$

$X_{u0}$= Min {the maximum size of bandwidth for upward direction in a LSP pair}

$X_d$= {the size of required bandwidth for downward direction}/$X_{d0}$

$X_{d0}$= Min {the maximum size of bandwidth for downward direction in a LSP pair}

For example, if there are two LSP pairs and the maximum size of bandwidth for upward direction in each LSP pair is 100Mb/s and 80Mb/s respectively, $X_{u0}$ will be 80Mb/s.

ii) Selection of one LSP pair

The LSP pair that satisfies the following three conditions will be selected:

- Min {the available size of bandwidth for the key direction in a LSP pair}
- Available upward bandwidth in the LSP pair is equal to or larger than the required upward bandwidth.
- Available downward bandwidth in the LSP pair is equal to or larger than the required downward bandwidth.

If there are two or more LSP pairs which satisfy the above three conditions, one LSP pair will be selected at random. Note that the request will be rejected if there are no LSP pairs that satisfy the above conditions.

iii) Allocation of bandwidth

Both upward and downward bandwidths are allocated simultaneously in the selected LSP pair.

**[Method C]**

As some services require a quick response, it would be necessary to consider network delay time when considering LSP pair selection. The network delay refers the time taken for a packet to be transmitted from one edge node to another edge node along the LSP pair. It is assumed here that each service request declares 'the permitted network delay time'. This method tries to select the LSP pair which has the maximum permissible network delay time as long as it is less than the maximum permitted network delay time, and to reserve as much bandwidth as possible for future requests which may require a short network delay time. This can minimize the request loss probability and reduce the total amount of bandwidth as a result.

Bandwidth allocation algorithm is as follows:

i) Selection of one LSP pair

The LSP pair that satisfies the following four conditions will be selected:

- Max {network delay time for the LSP pair}

- The network delay time for the LSP pair is less than the permitted network delay time of the request.

 - Available upward bandwidth in the LSP pair is equal to or larger than the required upward bandwidth.

 - Available downward bandwidth in the LSP pair is equal to or larger than the required downward bandwidth.

If there are two or more LSP pairs that satisfy the above four conditions, one LSP pair will be selected at random. Note that the request will be rejected if there are no LSP pairs that satisfy the above conditions.

ii) Allocation of bandwidth

Both upward and downward bandwidths are allocated simultaneously in the selected LSP pair.

## 4. Simulation evaluation

### 4.1 Simulation conditions

1) The evaluation is performed by a computer simulation using the C language.  The simulation model is based on the case where n is 2 in Figure 1. That is, there are two LSP pairs (LSP pair 1 and LSP pair 2) between edge node A and edge node B. The maximum size of upward bandwidth of LSP pair 1 and LSP pair 2 is assumed to be $U_1$ and $U_2$ respectively. The maximum size of downward bandwidth of LSP pair 1 and LSP pair 2 is assumed to be $D_1$ and $D_2$ respectively. Moreover, the network delay time of LSP pair 1 and LSP pair 2 is assumed to be $T_1$ and $T_2$ respectively.   Here, both $T_1$ and $T_2$ are assumed to be constant here.

When a service request is generated, LSP pair 1 or LSP pair 2 is selected according to the LSP selection method proposed in Section 3.

2) The size of required bandwidth for upward direction and that for downward direction follow a Gaussian distribution, and average values are given by $B_u$ and $B_d$ respectively.

3) It is supposed here that the following k requests will be generated repeatedly.   The value of $B_u$ and $B_d$ of the first generated request are $i_1$ and $j_1$ respectively.   The value of $B_u$ and $B_d$ of the second generated request are $i_2$ and $j_2$ respectively. The value of $B_u$ and $B_d$ of the k-th generated request are $i_k$ and $j_k$ respectively. We denote this request generation pattern like $\{B_u=i_1,B_d=j_1; B_u=i_2,B_d=j_2; \ldots; B_u=i_k,B_d=j_k\}$ in this paper.

4) The generation interval of requests follows an exponential distribution (average interval of request arrival is given by r). The service time H, which is the total time from a generation of the request to a completion of the service, is assumed to be constant. Each request occupies the allocated bandwidth until its service time H passes.

### 4.2 Simulation results and analysis

#### 4.2.1 Method A vs. Method B

Figures 3, 4, 5 and 6 illustrate the result of simulations which compare Method B with Method A. Figure 3(1) compares the request loss probability, which is the probability that either the bandwidth for upward direction or that for downward direction is not available, in the case where the value of $B_u$ is the same as that of $B_d$.  Figure 3(2) compares the request loss probability in the case where the sizes of bandwidth for the two directions ($B_u$ and $B_d$) rise and fall in anti-phase, that is a large bandwidth for the upward direction is followed by a large bandwidth for the downward direction. Figure 4 illustrates how much Method B can reduce the total bandwidth compared with Method A, while achieving the same request loss probability. $Z_b$ in Figure 4 shows how much Method B can reduce the total bandwidth compared with Method A to achieve the same request loss probability.  Figure 5 evaluates the impact of ratio of maximum bandwidth of each LSP pair on the request loss probability, assuming the total size of downward bandwidth ($D_1+D_2$) and that of upward bandwidth ($U_1+U_2$) are constant. Figure 6 evaluates the impact of the number of LSP pairs on the request loss probability with the same simulation conditions as those of Figure 3(2).

The following points are clear from these Figures:

(1) Method B performs better than Method A in the case where the sizes of bandwidth for two directions rise and fall in anti-phase. This is also true even if the number of LSP pairs increases except for the case where the number of LSP pairs is odd.

This is because the deadlock state as in Figure 2 occurs readily in Method B when the number of LSP pairs is odd and the sizes of bandwidth for two directions rise and fall in anti-phase. It is noted that the request loss probability will be small when the number of LSP pairs increases as in Figure 6, because the number of LSP pairs in the state that can be processed increases relatively for the same amount of service demands.

(2) Method B can reduce the total bandwidth required by up to 10% compared with Method A, as shown in Figure 4.

(3) To allocate bandwidth to one specific LSP pair as much as possible is better rather than to distribute bandwidth to multiple LSP pairs if the total bandwidth is constant.    For example, the request loss probability when $U_1$ and $D_1$ are 40 (the total bandwidth is allocated only to LSP pair 1) is small compared with the request loss probability when $U_1$ and $D_1$ are 20 in Figure 5.

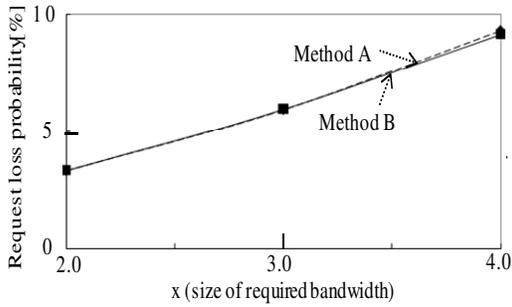

(1) $\{B_u=x, B_d=x\}$

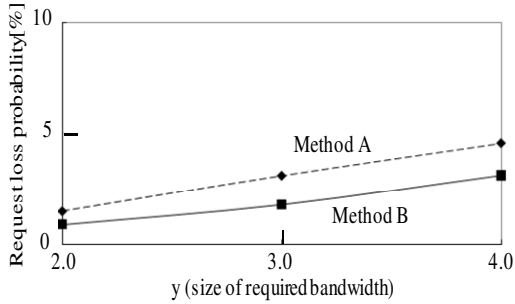

(2) $\{B_u=y, B_d=1; B_u=1, B_d=y\}$

$U_1=D_1=20, U_2=D_2=20, H=6$

Figure 3. Evaluation of request loss probability of Method A and Method B

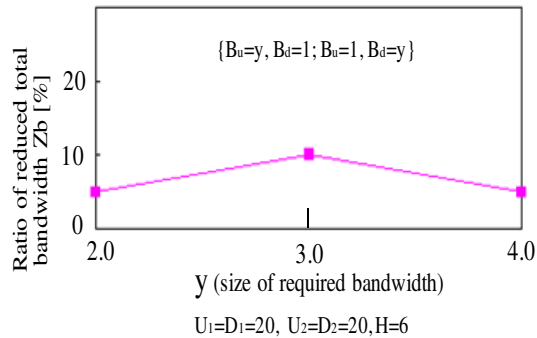

$U_1=D_1=20, U_2=D_2=20, H=6$

Figure 4. Evaluation of reduced bandwidth by Method

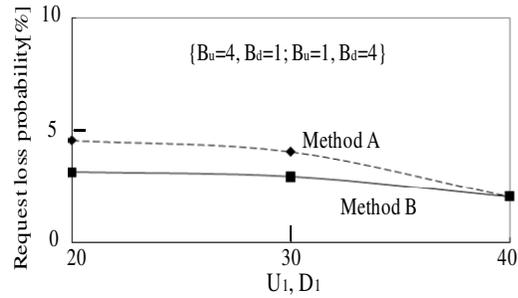

$U_1+U_2=40, D_1+D_2=40, H=6$

Figure 5. Evaluation of size of bandwidth when total amount of bandwidth is constant

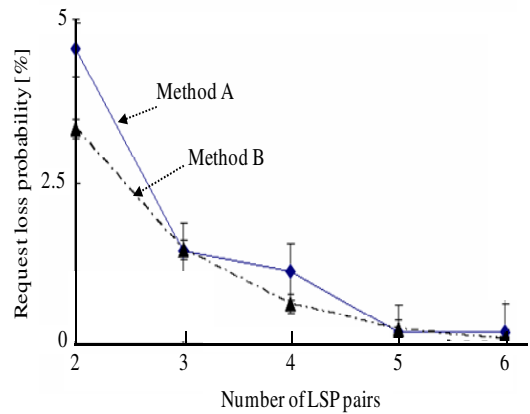

$U_1=U_2=20, D_1=D_2=20, H=6; \{B_u=4, B_d=1; B_u=1, B_d=4\}$

Figure 6. Evaluation of the number of LSP pairs

## 4.2.2 Method B vs. Method C

Figure 7, in which network delay time is considered, illustrates the result of simulations which compare Method C with Method B. Figure 7(1) evaluates the impact of ratio S of number of requests with short permitted delay time to total requests on request loss probability, and Figure 7(2) evaluates the impact of ratio S on the amount of total bandwidth reduction by Method C. The maximum permitted network delay time for each request is assumed to be either 0.1 [sec] or 0.3 [sec]. It is also assumed that $T_1$ and $T_2$ are 0.1 [sec] and 0.3[sec] respectively. Figure 7 assumes the case where the sizes of bandwidth for two directions rise and fall in anti-phase. The vertical axis, $Z_c$, in Figure 7(2) shows how much Method C can reduce the total bandwidth compared with Method B to achieve the same request loss probability. For example, if Zc is 10%, Method C can reduce 10% of total bandwidth compared with Method B.

The following points are clear from Figure 7:

(1)The request loss probability of Method C is smaller than that of Method B, regardless of the value of ratio S.

This is because both LSP pair 1 and LSP pair 2 can be selected for all requests when the network delay time is not considered in the case of LSP pair selection, but only LSP pair 1 can be selected for requests which require a short network delay time when the network delay time is considered in the case of LSP pair selection. In a word, the restriction of network delay time is more severe than that of the size of bandwidth.

It is noted that the difference of request loss probability between Method B and Method C reaches its maximum when ratio S is around 70%. This is because the request loss will easily occur in method B when the bandwidth of LSP pair 1 are used by many requests with short

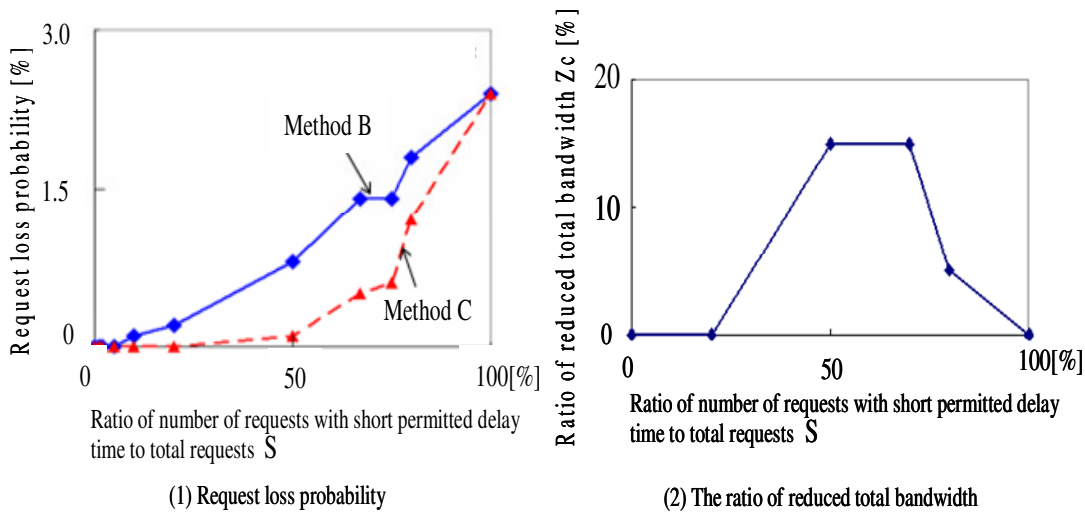

(1) Request loss probability　　　　　　(2) The ratio of reduced total bandwidth

{$B_u$=4,$B_d$ =2;$B_u$=2,$B_d$=4}, $U_1$=$D_1$=20,$U_2$=$D_2$ =20, H=6, $T_1$=0.1sec, $T_2$=0.3sec

Figure 7. Comparative evaluation of Method B and Method C

permitted network delay time and a request with short permitted network delay time is generated just after a request with long permitted network delay time is allocated to LSP pair 1. This will occur frequently S is around 70%.

(2) Method C can reduce the total bandwidth required by up to 15% compared with Method B, as illustrated in Figure 7(2). It is also noted that Zc becomes maximum when S is around 70%. This is the same reason as the above (1).

# 5. Conclusions

This paper has proposed two new methods of selecting the optimal LSP pair in MPLS networks, on the assumption that both upward and downward LSPs are established as a pair (both-way operation) and multiple LSP pairs are established between the same pair of edge nodes. It is supposed that the optimal LSP pair is selected from among multiple LSP pairs for each service request and both upward and downward bandwidths are allocated simultaneously in the selected

LSP pair.

It has been demonstrated by simulation evaluation that the proposed methods could reduce the total bandwidth required by up to 15%, compared with the conventional method. As the design and management of many of existing networks are based on both-way operation, the proposed methods can also become possible to reuse the know-how and management tools in the existing network. It should be noted that the method proposed here can also be applied to LSP selection in GMPLS (Generalized MPLS) networks or the link selection in link aggregation configuration of Ethernet.

## Acknowledgment

We would like to thank Mr. Shigehiro Tsumura and Mr. Kenichi Hatakeyama for their help with the simulation.

## Author

**Shin-ichi Kuribayashi**    received the B.E., M.E., and D.E. degrees from Tohoku University, Japan, in 1978, 1980, and 1988 respectively. He joined NTT Electrical Communications Labs in 1980.   He has been engaged in the design and development of DDX and ISDN packet switching, ATM, PHS, and IMT 2000 and IP-VPN systems. 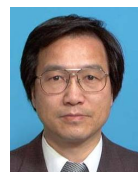 He researched distributed communication systems at Stanford University from December 1988 through December 1989. He participated in international standardization on ATM signaling and IMT2000 signaling protocols at ITU-T SG11 from 1990 through 2000.   Since April 2004, he has been a Professor in the Department of Computer and Information Science, Faculty of Science and Technology,   Seikei University.   He is a member of IEEE, IEICE and IPSJ.